\documentclass[sigconf,nonacm]{acmart}
\usepackage{algorithm}
\usepackage{algorithmic}
\usepackage{enumitem}
\usepackage{pifont}
\newcommand{\cmark}{\ding{51}}%
\newcommand{\xmark}{\ding{55}}%

\AtBeginDocument{%
  \providecommand\BibTeX{{%
    \normalfont B\kern-0.5em{\scshape i\kern-0.25em b}\kern-0.8em\TeX}}}

\setcopyright{rightsretained}

\usepackage{algorithm}
\usepackage{algorithmic}
\usepackage{breakcites,balance}

\begin{document}

\title{Blockwise Feature Interaction in Recommendation Systems}

\author{\vspace{0.2in} Weijie Zhao, \ Ping Li\vspace{0.05in}}
\affiliation{\institution{LinkedIn Ads\vspace{0.05in}\\
700 Bellevue Way NE, Bellevue,  WA 98004, USA\\
\{zhaoweijie12, pingli98\}@gmail.com \vspace{0.2in}}\country{\vspace{0.3in}}}

\begin{abstract}
Feature interactions can play a crucial role in recommendation systems as they capture complex relationships between user preferences and item characteristics. Existing methods such as Deep \& Cross Network (DCNv2) may suffer from high computational requirements due to their cross-layer operations. In this paper, we propose a novel approach called \textbf{blockwise feature interaction (BFI)} to help alleviate this issue. By partitioning the feature interaction process into smaller blocks, we can significantly reduce both the memory footprint and the computational burden. Four variants (denoted by \textbf{P}, \textbf{Q}, \textbf{T}, \textbf{S}, respectively) of BFI have been developed and empirically compared. Our experimental results demonstrate that the proposed algorithms achieves close accuracy  compared to the standard DCNv2, while greatly reducing the computational overhead and the number of parameters. This paper contributes to the development of efficient recommendation systems by providing a practical solution for improving feature interaction efficiency.
\end{abstract}

\maketitle


\section{Introduction}\label{sec:intro}
In the recent decade, deep learning has emerged as a powerful technique for building recommendation systems that can effectively capture complex patterns and user preferences~\cite{karpathy2014deep,covington2016deep,yu2018hierarchical,fan2019mobius, zhao2019aibox, chang2020pre,li2020video, giorgi2021declutr,yu2022egm,spillo2022knowledge,lanchantin2021general,fei2021gemnn, neelakantan2022text}. One significant challenge in designing these systems lies in capturing feature interactions, which are essential for achieving high performance and accuracy~\cite{rendle2010factorization,guo2017deepfm,lian2018xdeepfm,wang2021dcnv2}.

\textbf{Feature interactions} refer to the relationships and dependencies between different input features, such as user demographics, historical preferences, and item characteristics. These interactions can be highly non-linear and intricate, making it challenging to model them accurately using traditional linear models or shallow neural networks. The number of potential feature interactions grows exponentially with the number of features, making it infeasible to explicitly enumerate and represent all possible interactions. Even if we only consider second-order feature interactions, we will still have to explicitly introduce a quadratic number of combined features. As the number of features increases, the computational and memory requirements become unpractical.

\begin{figure*}[t]
\centering 
\mbox{
\includegraphics[width=2.2in]{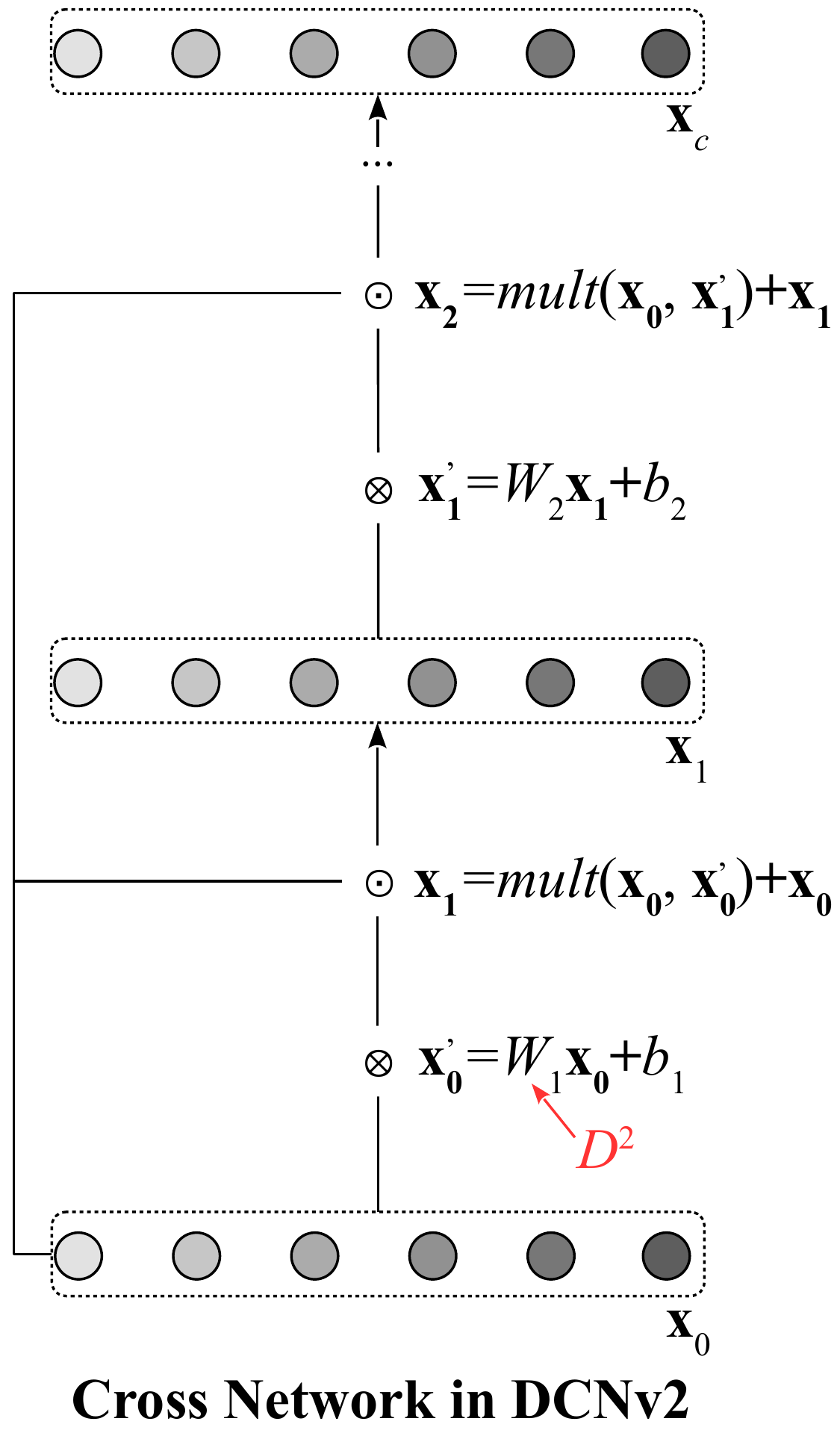}
\hspace*{0.6in}
\includegraphics[width=2.88in]{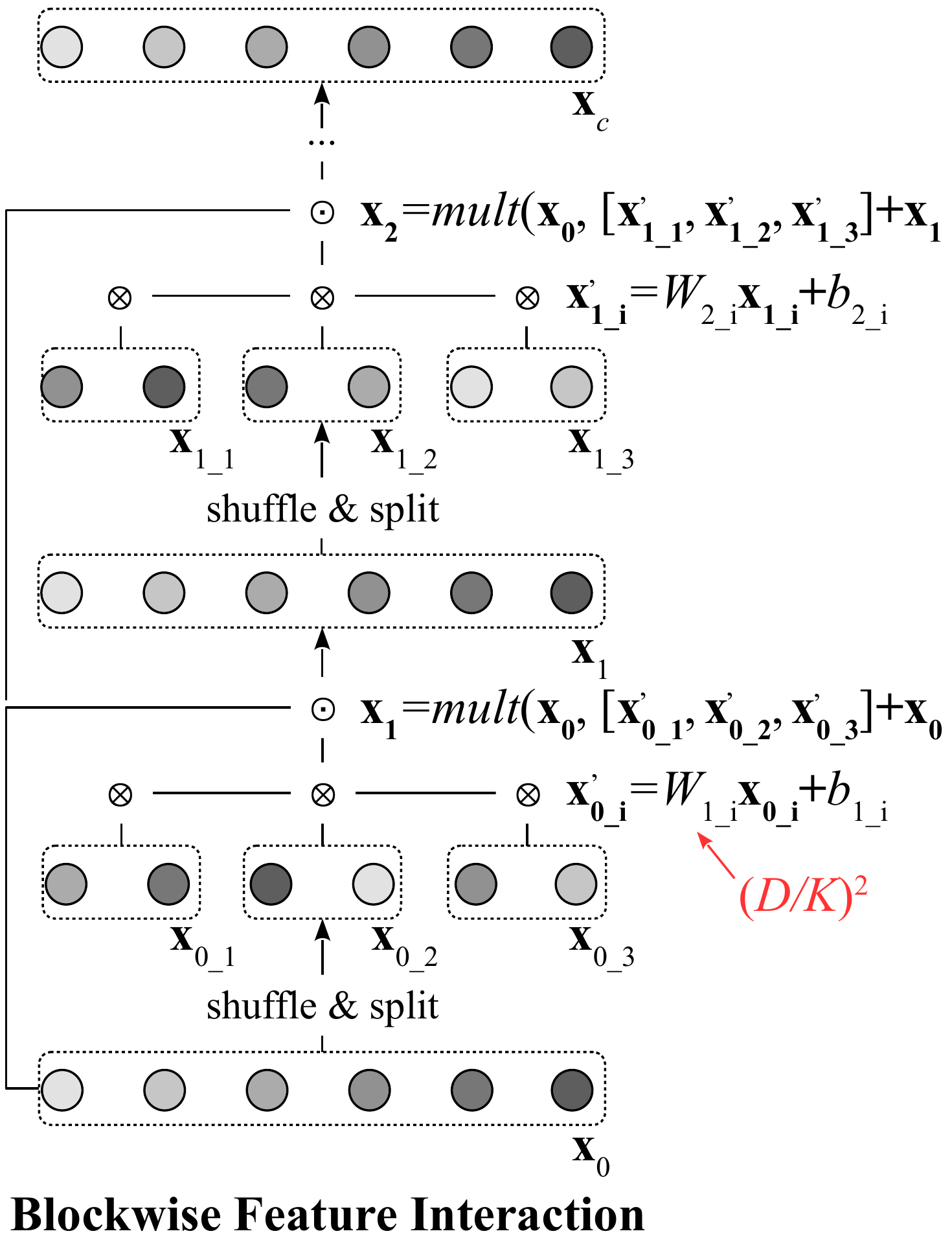}
}

\caption{A visual illustration of the difference between the cross network in DCNv2 and our proposed Blockwise Feature Interaction (BFI) architecture, where mult is the element-wise multiplication, [] is the concatenation operator, D is the dimension of the input vector, and K is the block factor in BFI (K=3 in this example).}\label{fig:example}
\end{figure*}

\textbf{Feature interactions in deep learning.} Deep learning, with its ability to automatically learn hierarchical representations and complex relationships, has shown great promise in addressing this challenge. For example, Deep \& Cross Network (DCN) and its improved version DCNv2~\cite{wang2021dcnv2} demonstrate remarkable performance in capturing feature interactions. DCNv2 combines the strengths of deep neural networks and cross-networks, allowing for the learning of both low- and high-order feature interactions. We show an example of cross networks in the left part of Figure~\ref{fig:example}. By leveraging cross-network architectures, DCNv2 enables the model to explicitly model the interactions between features, enhancing the overall predictive power of the system.  

\textbf{Cost of cross networks in DCNv2.} Despite its effectiveness, the cost associated with DCNv2 cannot be neglected. As shown in the left part of Figure~\ref{fig:example}, the complexity of the cross networks in DCNv2 grows quadratically ($D^2$) with the number of input features ($D$), resulting in a substantial computational burden and increased training time. This cost factor is particularly pronounced in large-scale recommendation systems that handle vast amounts of features and require real-time or near-real-time predictions. As a result, mitigating the cost while maintaining the desired level of performance becomes a critical consideration for deploying DCNv2-based recommendation systems.

\textbf{Blockwise Feature Interaction (BFI).} 
In this paper, we aim to explore the trade-off between performance and cost of the cross network in DCNv2 to alleviate the quadratic factor. 
We propose Blockwise Feature Interaction (BFI), a network architecture that partitions the cross layers into smaller blocks. By splitting the vectors for the fully-connected embedding layers, the $D^2$ cost is reduced to $(\frac{D}{K})^2 \times K = \frac{D^2}{K}$, where D represents the total number of input features, and K is the number of blocks in BFI. The right part of Figure~\ref{fig:example} depicts the architecture of BFI when $K=3$.

\vspace{0.1in}
\textbf{Contributions.}
The main contributions of this paper are:
\begin{itemize}[leftmargin=*]
\item We introduce Blockwise Feature Interaction (BFI), that reduces both computation and memory consumption of cross networks for modeling feature interactions in recommendation systems.
\item We propose four variants of BFI which are designed for saving the memory and evaluation cost, by reducing the number of parameter shuffling and introducing parameter sharing.
\item We experimentally evaluate the proposed methods on public datasets. The empirical results confirm the effectiveness of our proposed methods.
\end{itemize}

\section{Blockwise Feature Interaction}

\subsection{Cross Network}
As shown in the left part of Figure~\ref{fig:example}, the cross network operates by taking the outputs of the input layer and feeding them into a series of cross layers. Each cross layer captures a specific level of interaction between the features. The cross layers consist of a combination of linear and nonlinear operations that enable the model to learn complex feature interactions. 
The key mechanism in the cross network is the cross operation, which calculates the element-wise product of different features and applies a weight ($W$) to each interaction. This weight reflects the importance or contribution of the specific interaction to the overall prediction. By incorporating these weighted feature interactions, the cross network enhances the expressive power of the model, enabling it to capture complex patterns and dependencies that might not be evident in individual features.

During the training process, the cross network learns the weights associated with the feature interactions through backpropagation and gradient descent optimization. The model adjusts the weights to minimize the discrepancy between its predictions and the ground truth labels. This iterative learning process allows the cross network to fine-tune its parameters and improve its ability to capture relevant feature interactions.

The cross network can capture both low-order and high-order feature interactions. Low-order interactions refer to simple pairwise interactions between two features, while high-order interactions involve combinations of three or more features. By incorporating $C$ cross layers, the cross network can capture increasingly complex and higher-order interactions.
The larger $C$ is, the higher-order feature interaction is captured. 

\textbf{Cost of cross network.} The cost of feeding forward an instance with $D$ features through an $C$-layer cross network is $O(D^2 \cdot C)$---the $1 \times D$ vector and $D \times D$ matrix multiplication for each layers. The space complexity is $O(D^2) \cdot C$ for $C$ weight matrices. Both of them are quadratic to the number of features, making it inefficient or even infeasible to process in latency-limited real-world systems and to store the feature interaction weights.

\subsection{Cross Weight Partitioning}
To address the quadratic computational cost associated with capturing feature interactions in recommendation systems, we propose the Blockwise Feature Interaction (BFI) method. BFI aims to reduce the computational overhead while maintaining or even improving the performance of recommendation systems.

The quadratic term in the computational and memory costs arises from the weight matrices $W$, where the size of $W$ is $D \times D$. However, by partitioning the features into $K$ parts, we can effectively reduce the size of $W$ to $(D/K) \times (D/K)$, resulting in smaller matrix sizes of $(D/K)^2$. Despite having $K$ matrices due to the partitioning (one matrix for each partition), the total size of these $K$ matrices is $D^2/K$, which is $K$ times smaller than the original cross network.

This partitioning strategy offers a practical solution to mitigate the quadratic time and memory costs associated with the cross network, enabling the efficient capture of feature interactions in recommendation systems.

\newpage

\begin{algorithm}[htbp]
\caption{Blockwise Feature Interaction Network}\label{alg:bfi}
\begin{flushleft}
\textbf{Input:} Feature vector $x_0$, cross network layers $C$, BFI factor $K$.\\
\textbf{Output:} Embedded feature vector $x_{C}$
\end{flushleft}
\begin{algorithmic}[1]
\FOR{$c$ $\leftarrow$ $1$ \textbf{to} $C$}
    \STATE $x_{c\_1},x_{c\_2}\cdots x_{c\_K}$ $\leftarrow$ shuffle\&split($x_{c-1}$, $K$)
    \FOR{$i$ $\leftarrow$ $1$ \textbf{to} $K$}
        \STATE $x_{c-1\_i}'$ $\leftarrow$ $W_{c\_i}x_{c-1\_i} + b_{c\_i}$
    \ENDFOR
    \STATE $x_{c}$ $\leftarrow$ mult($x_0$, $\left[x_{c-1\_1}', x_{c-1\_2}', \cdots x_{c-1\_K}'\right]$) + $x_{c-1}$
\ENDFOR
\RETURN $x_{C}$
\end{algorithmic}
\end{algorithm}

Algorithm~\ref{alg:bfi} outlines the workflow of our proposed BFI method. The algorithm takes a feature vector, $x_0$, as input along with the number of cross network layers, $C$, and the BFI factor, $K$. It aims to produce an embedded feature vector, $x_{C}$, which represents the input features with captured interactions. The algorithm begins by initializing a loop over the cross network layers, from $c=1$ to $C$. Within each layer, the feature vector $x_{c-1}$ from the previous layer is shuffled and split into $K$ parts. The purpose of this step is to partition the features into smaller blocks to reduce computational costs.
After that, we go through an embedding layer for each partitioned feature and merge them as the $D$-dimensional embedding. The embedding is then processed as normal cross network layers: elementwise multiplication and addition operations.

\subsection{Shuffle \& Split}
Note that the ``shuffle \& split'' operation mentioned in the algorithm serves two important purposes in capturing feature interactions within the Blockwise Feature Interaction Network.

\begin{figure*}[ht]
\includegraphics[width=.24\textwidth]{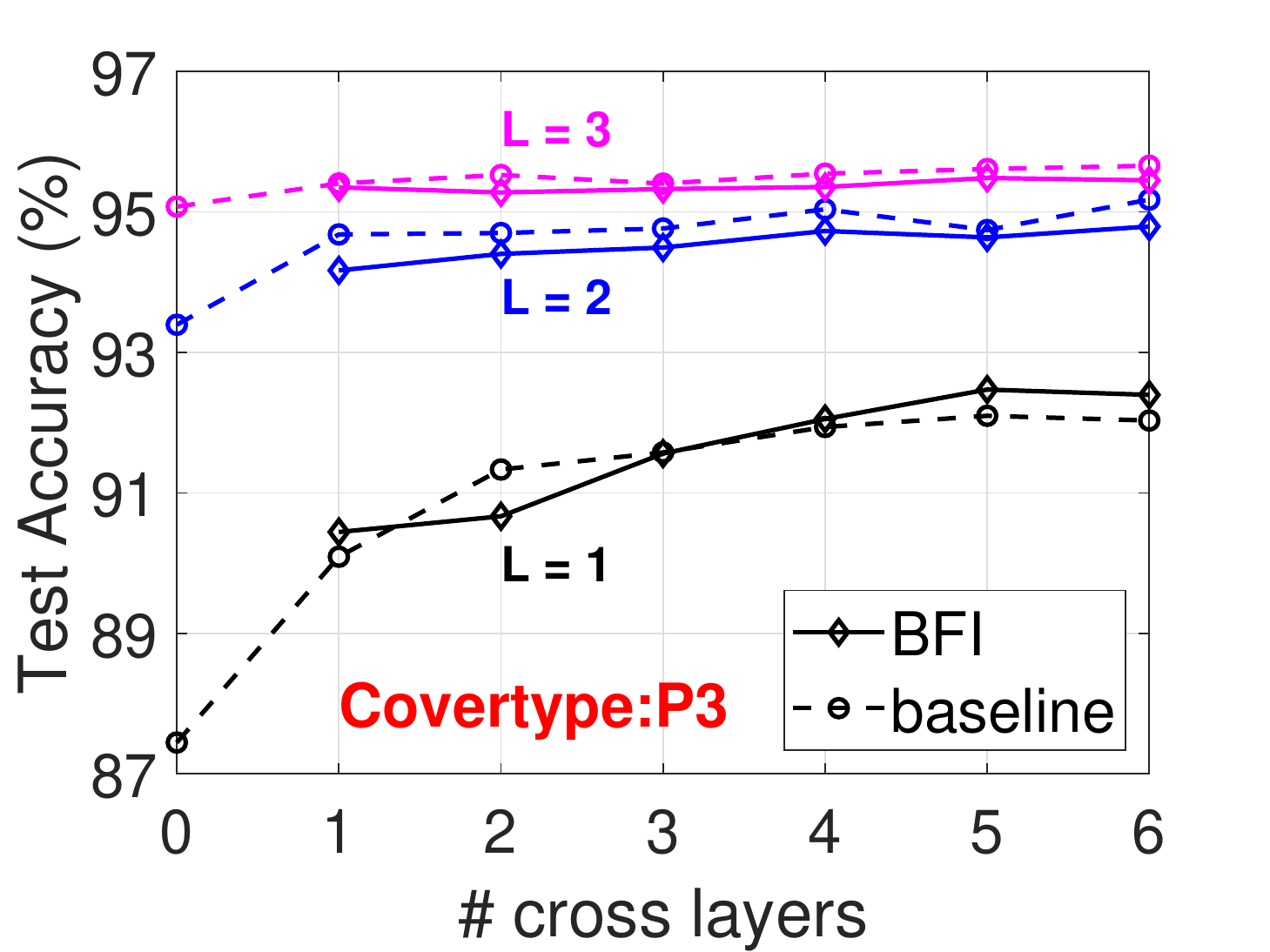}
\includegraphics[width=.24\textwidth]{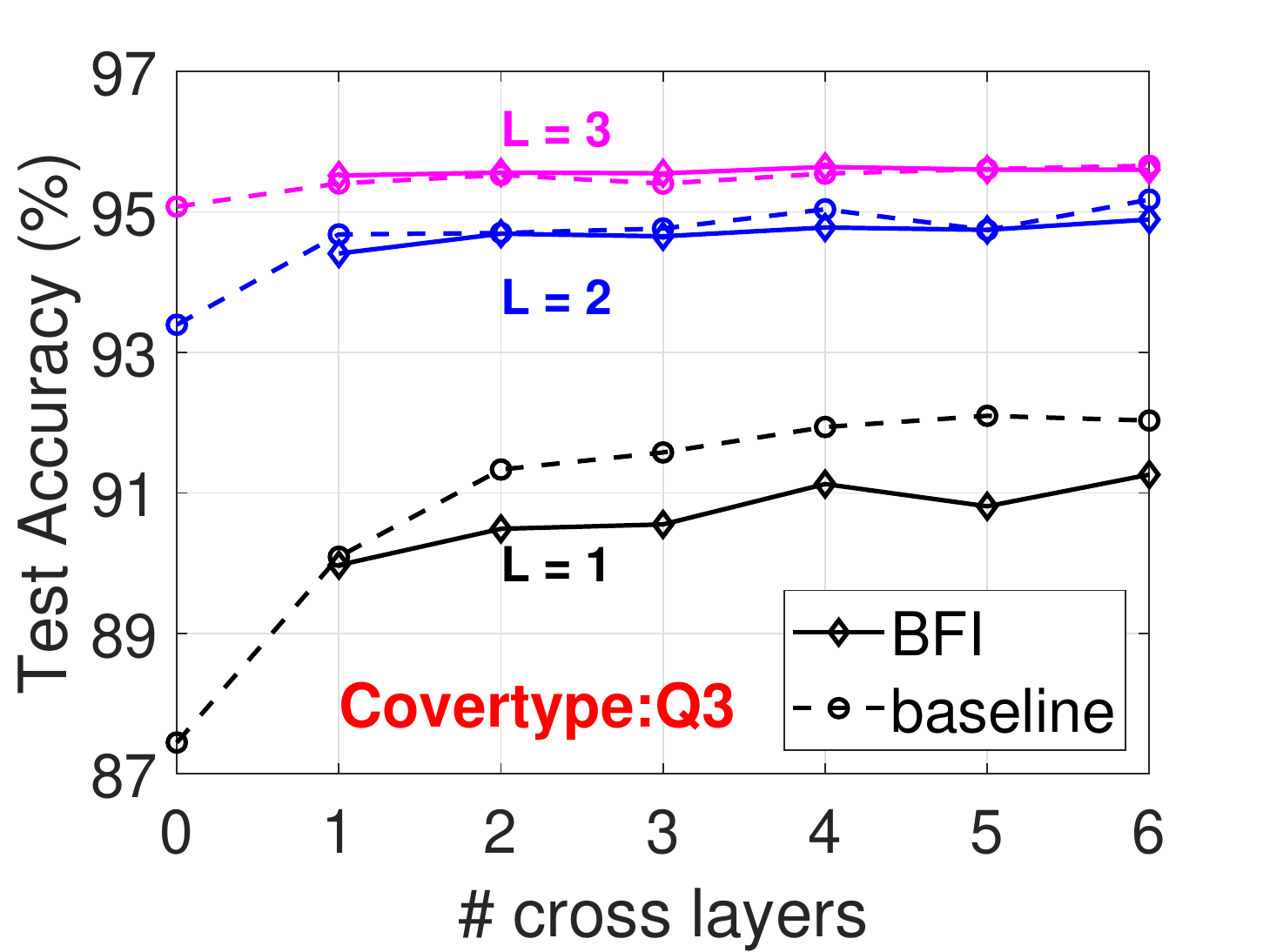}
\includegraphics[width=.24\textwidth]{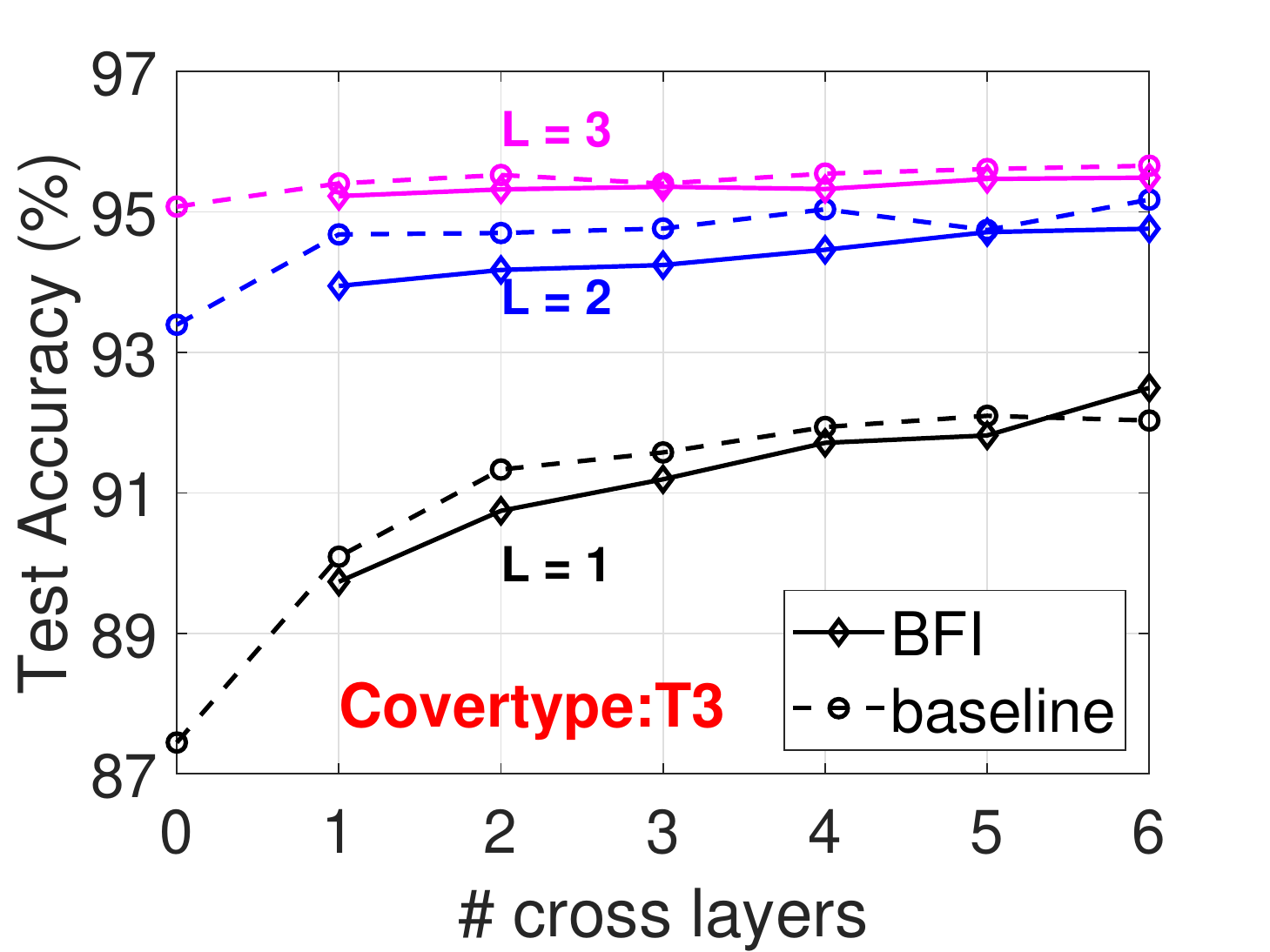}
\includegraphics[width=.24\textwidth]{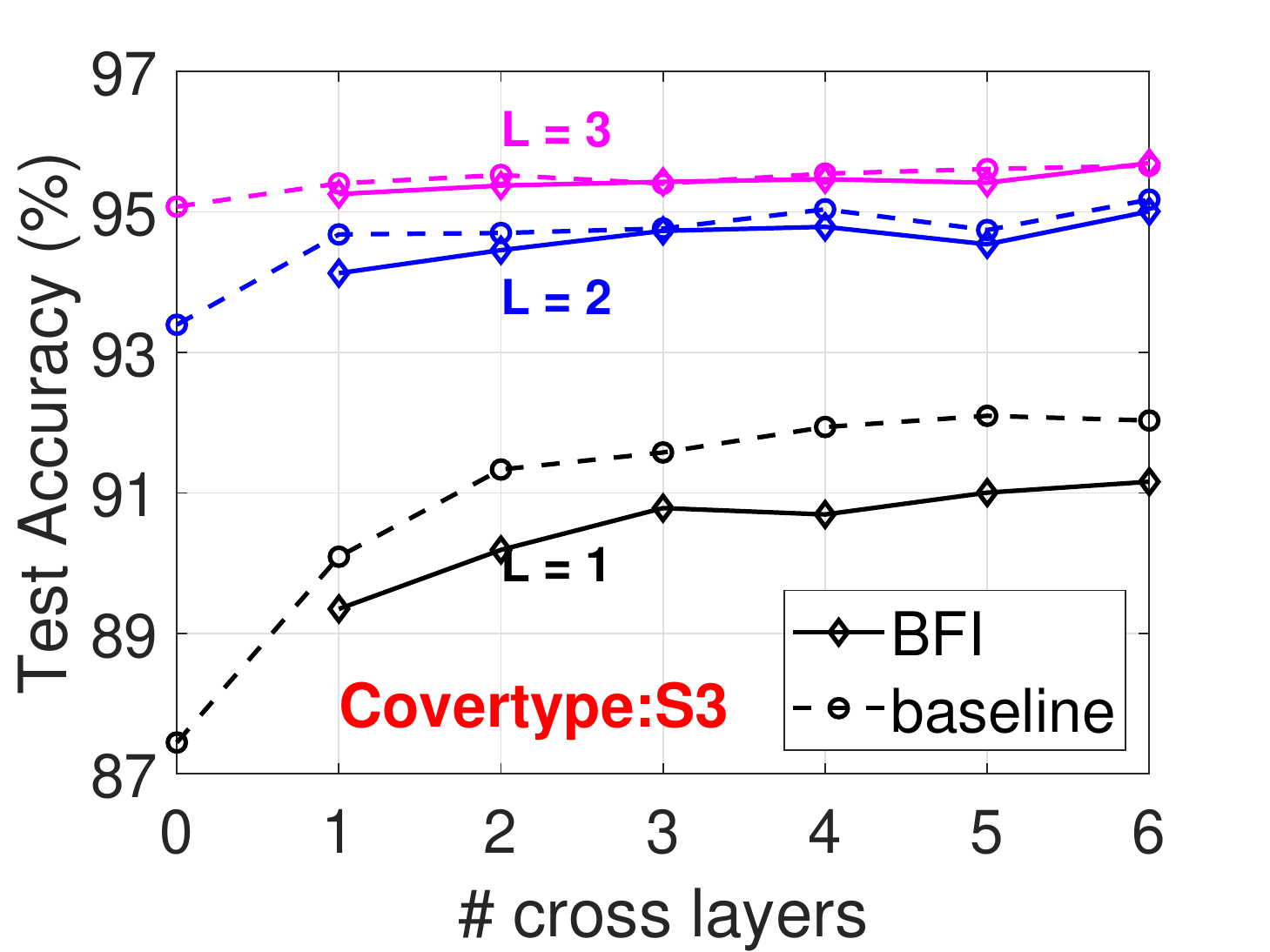}\\
\includegraphics[width=.24\textwidth]{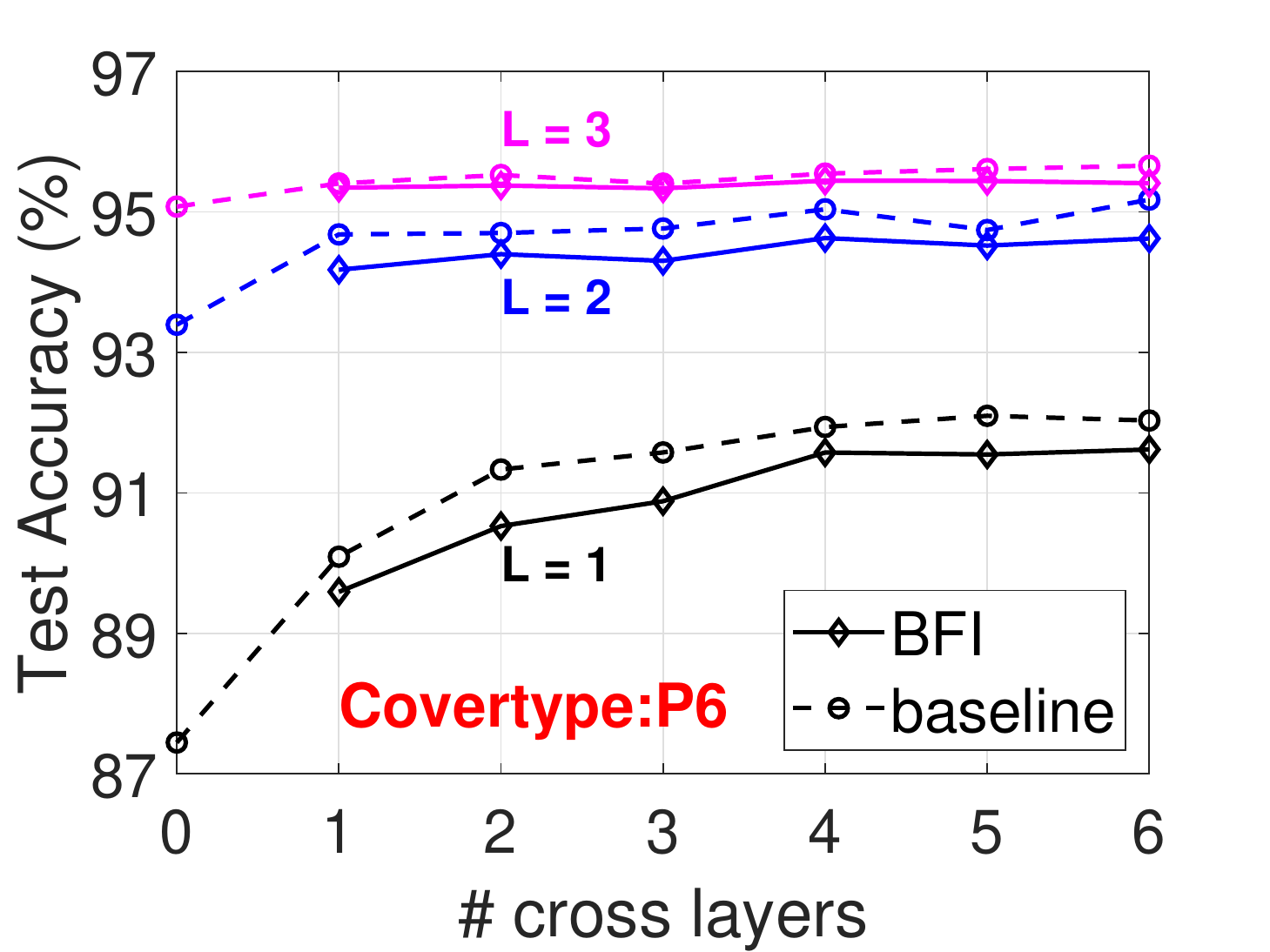}
\includegraphics[width=.24\textwidth]{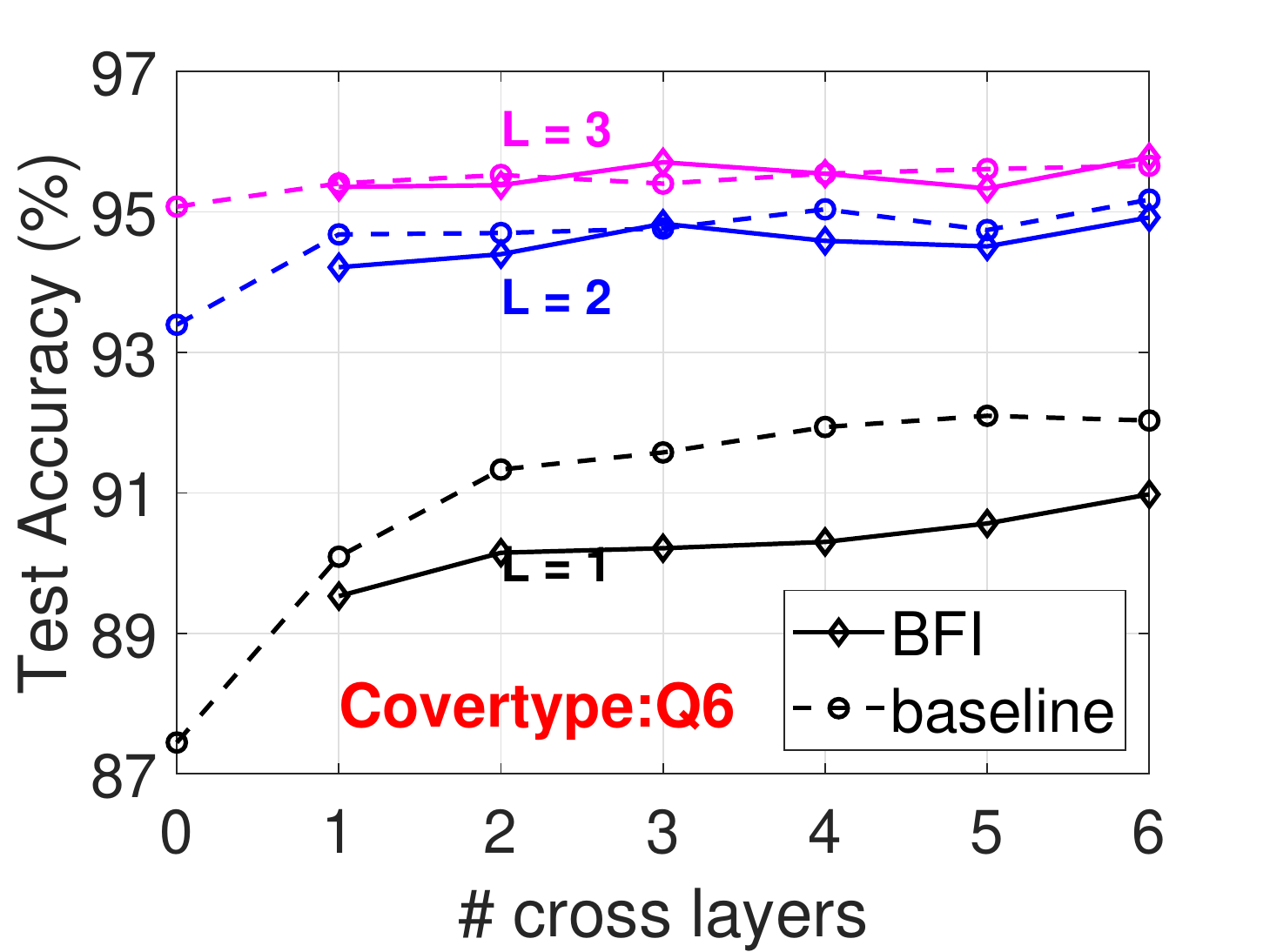}
\includegraphics[width=.24\textwidth]{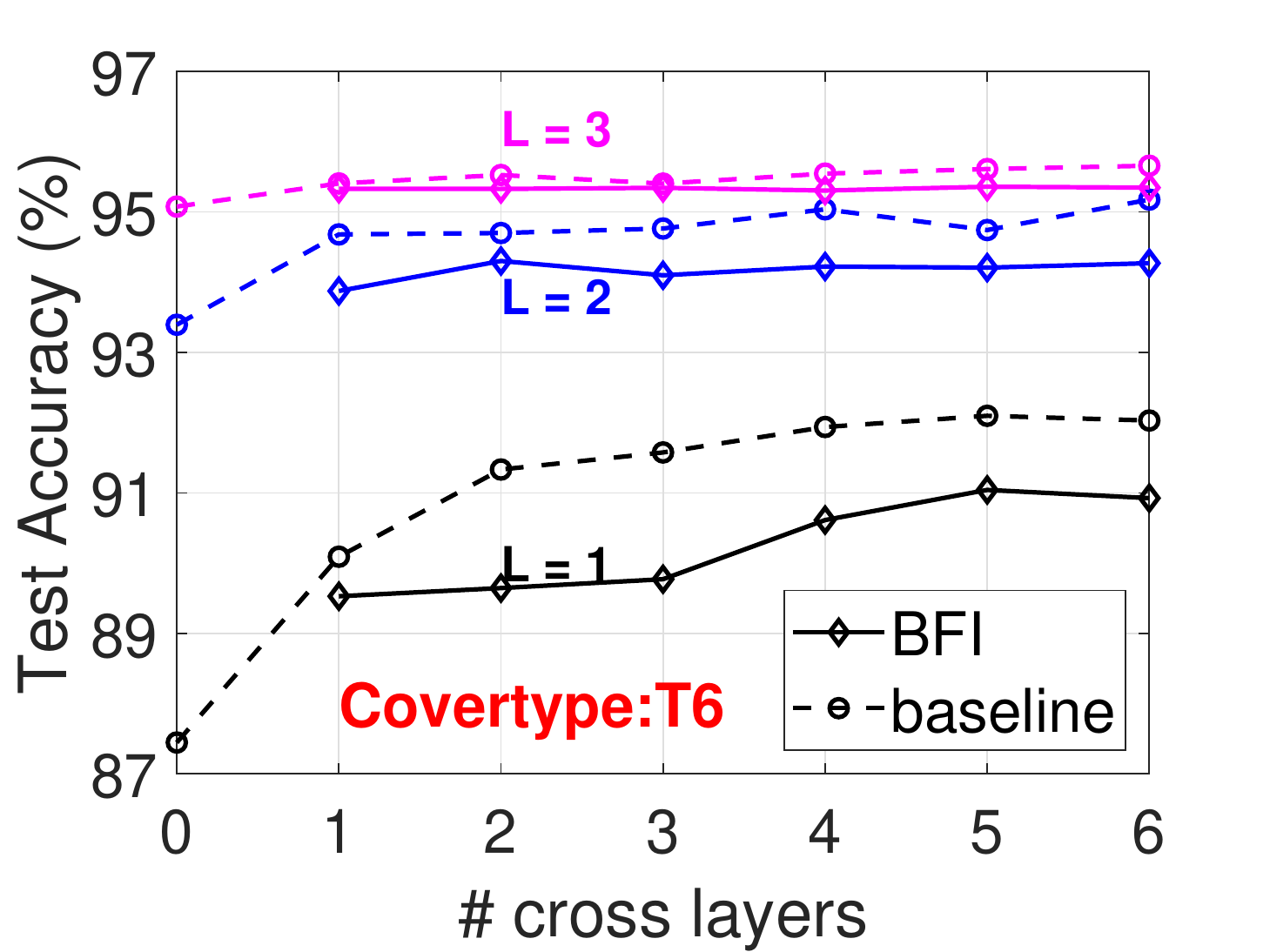}
\includegraphics[width=.24\textwidth]{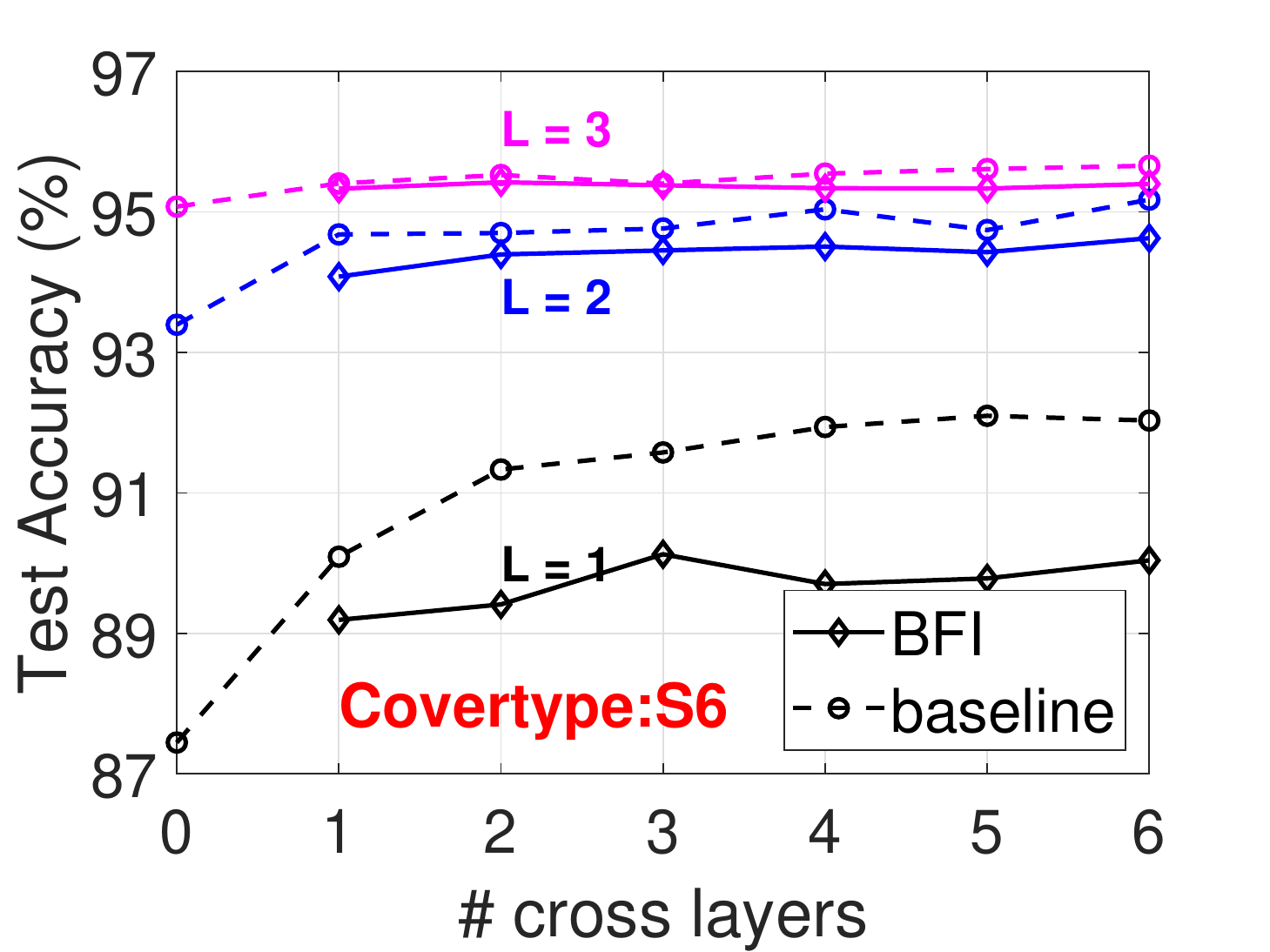}\\
\includegraphics[width=.24\textwidth]{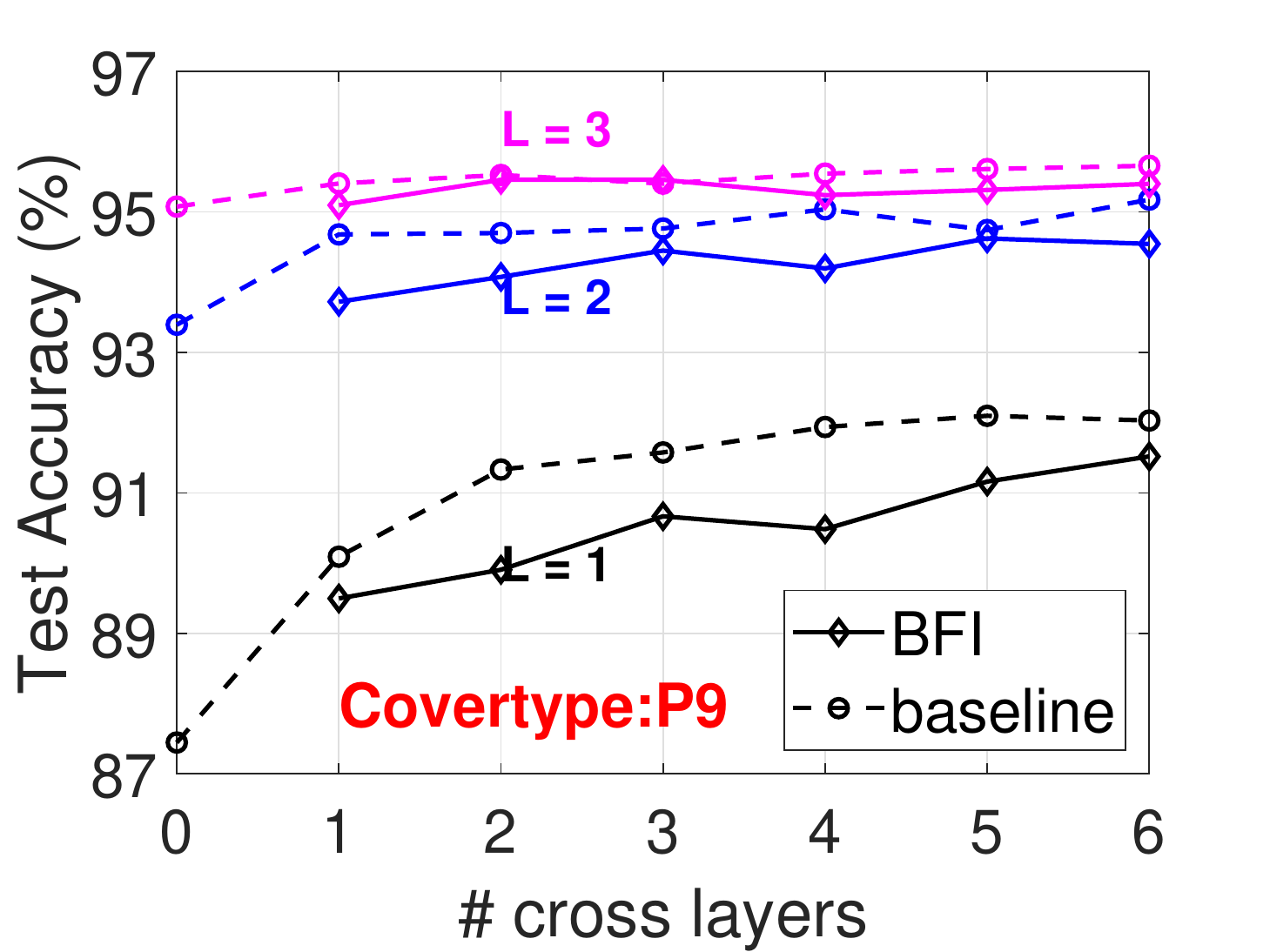}
\includegraphics[width=.24\textwidth]{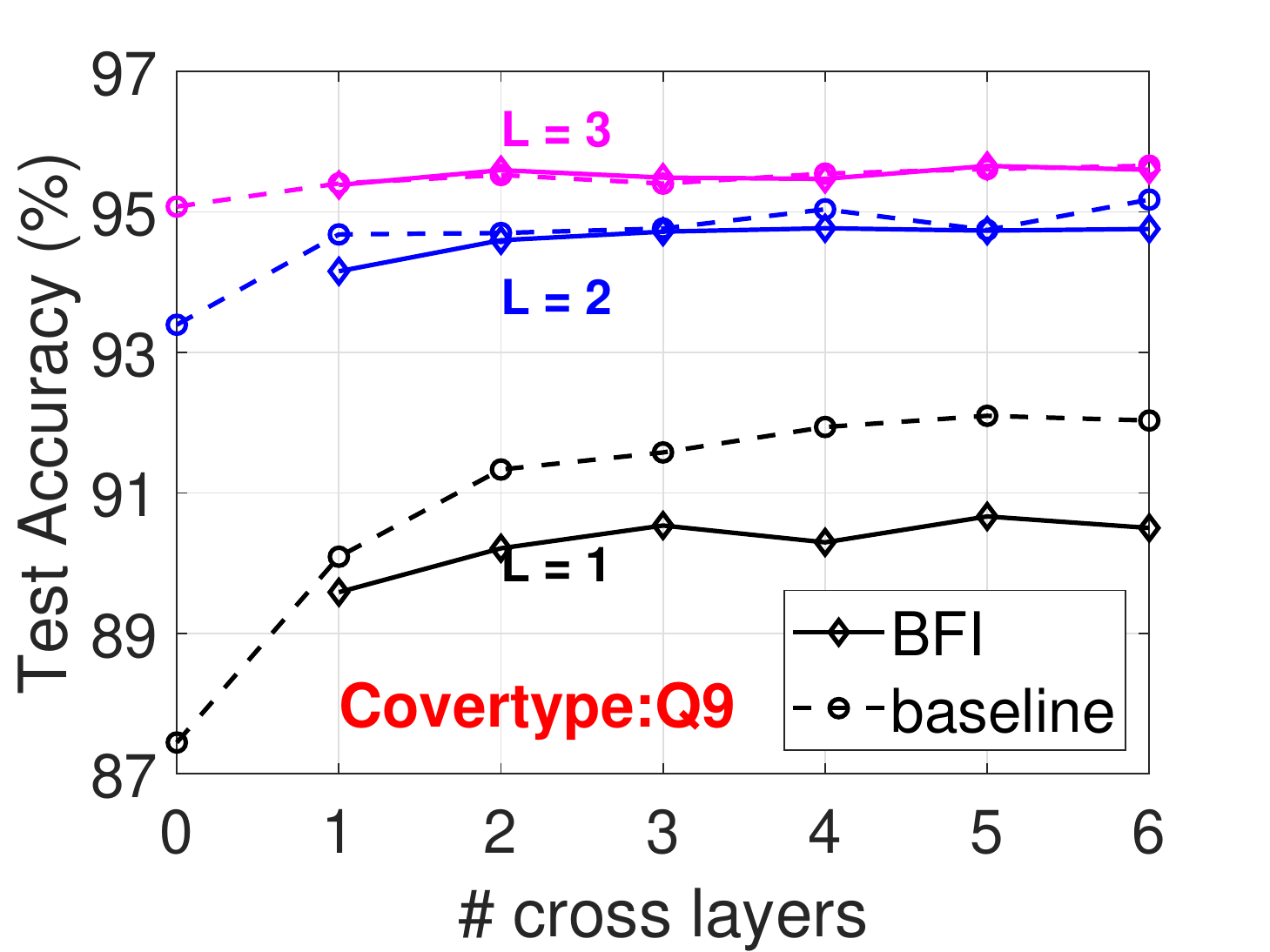}
\includegraphics[width=.24\textwidth]{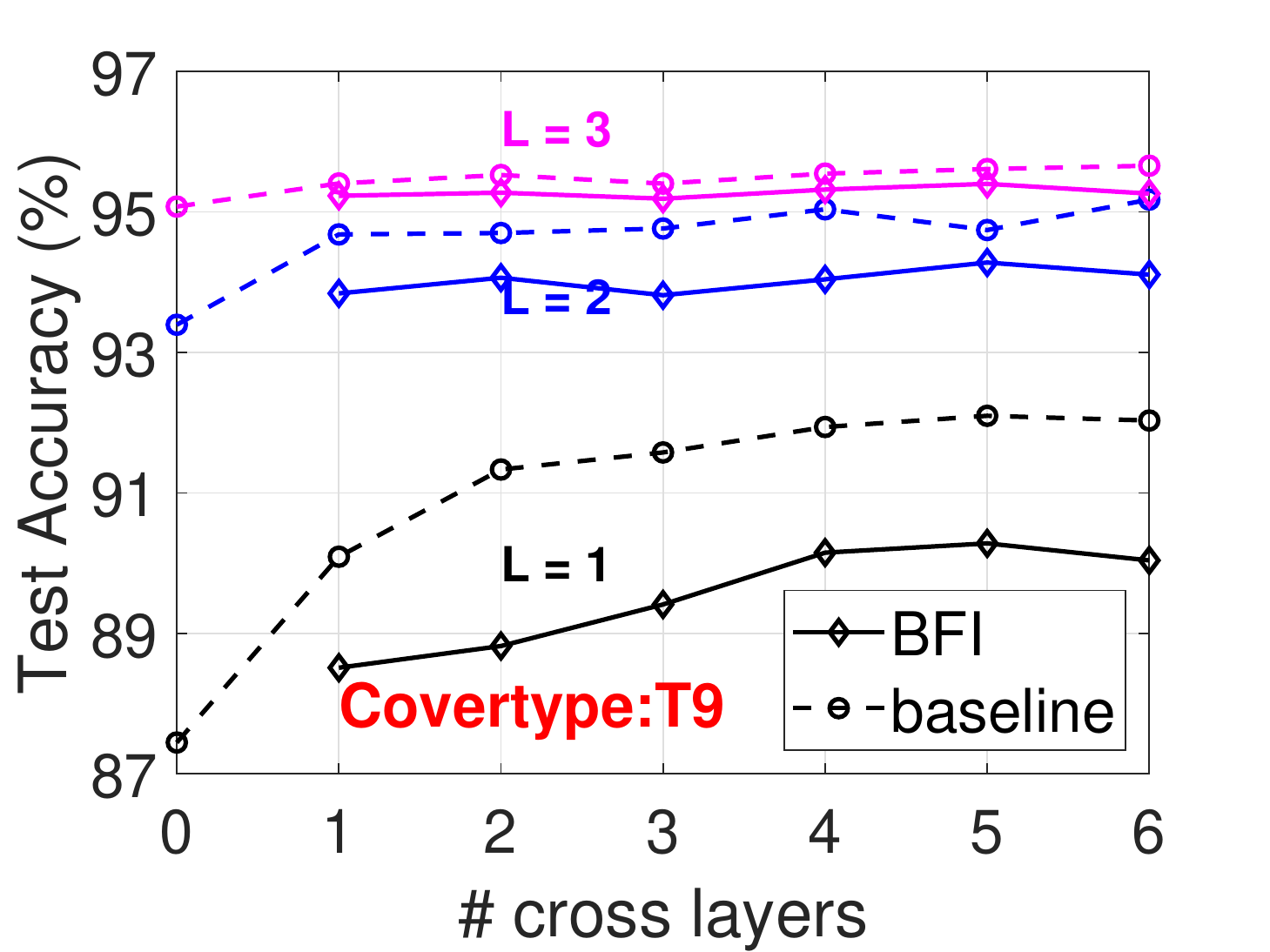}
\includegraphics[width=.24\textwidth]{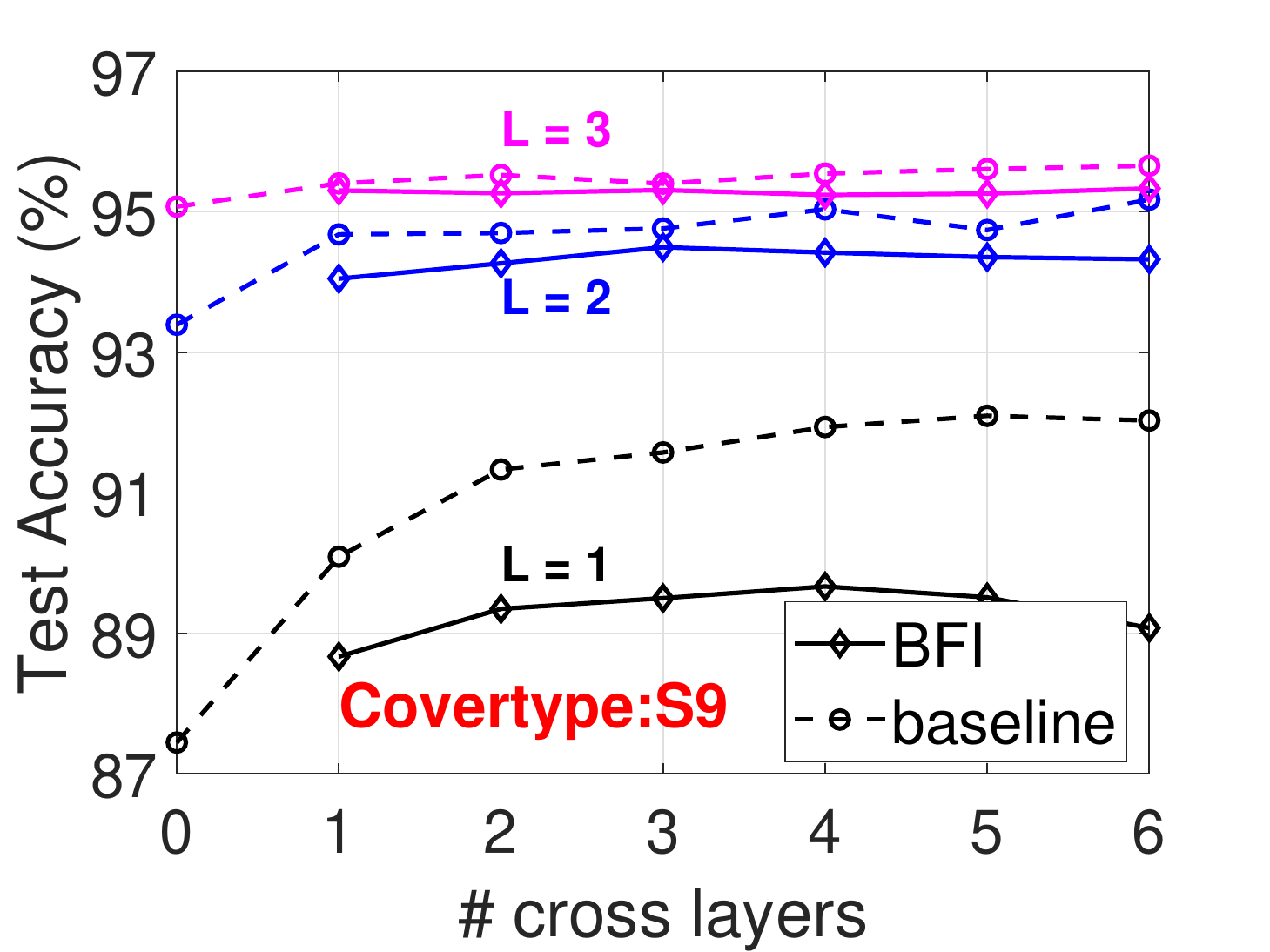}
\caption{Test accuracy for BFI and the cross network in DCNv2 (baseline) on Covertype.}\label{fig:covtype}
\end{figure*}

\begin{enumerate}
\item \textit{Addressing Dependencies in Feature Indices}: 
The shuffling step is necessary because the original feature indices may not be independent. For example, if the feature indices are continuous and represent similar characteristics, it implies that adjacent features may have higher chances of interacting with each other. By shuffling the feature vector, the algorithm introduces randomness and breaks any potential ordering or dependencies among the features. This ensures that the subsequent splitting does not bias the interactions solely based on their original order.

\item \textit{Enabling Interactions Across Blocks}:
Without shuffling, each partition would be independent across multiple cross layers. This means that the features within different blocks would have no direct interaction with each other, limiting the network's capacity to capture comprehensive feature dependencies. By shuffling the feature vector, the algorithm introduces randomness in the arrangement of the features, allowing for cross-block interactions to occur. As a result, the subsequent processing of each partition in the cross network layers can capture and model feature interactions both within and across the blocks, enhancing the network's ability to learn complex dependencies.
\end{enumerate}


\subsection{Interaction Weight Sharing}
After applying BFI, the memory required for storing the feature interaction weights is reduced from $O(D^2)$ to $O(D^2/K)$. We find that there is a potential to further reduce the memory footprint by making all partitions share a single embedding.

The observation is that during the embedding process, the feature interaction weights take an embedding as input and generate new embeddings for higher-order interactions. The only difference across the partitions is that they use embeddings from different partitioned blocks. Therefore, we can eliminate the need for separate embeddings per partition and instead share a single embedding across all partitions.

By sharing one embedding across all partitions, the memory requirement can be substantially reduced to $O(D^2/K^2)$. This reduction is significant because when training and inferencing deep learning models, the computations are often performed on GPUs, which typically have limited memory capacity. The difference in memory consumption by a factor of $K$ can be critical in determining whether the model fits within the available GPU memory or not.

\begin{table}[htbp]
\caption{Legend explanation.}\label{tbl:legend}
\vspace*{-.1in}
\begin{tabular}{ccc}
\toprule
\midrule
Legend & Shuffle & Share\\
\midrule
P & every layer & \xmark\\
Q & only once & \xmark\\
T & every layer & \cmark\\
S & only once & \cmark\\
\midrule
\bottomrule
\end{tabular}
\end{table}

In Section~\ref{sec:exp}, 
we enumerate the following 2 options:  (1) whether shuffling at each cross layer to address cross block interaction or only shuffling once on the original feature to address the feature indices dependencies; (2) whether enabling weight sharing. The 4 variants are denoted by P, Q, T, and S, respectively (Table~\ref{tbl:legend}).

\section{Experimental Evaluation}\label{sec:exp}
The objective of the experimental evaluation is to investigate the overall performance, as well as the impact of parameters, of the proposed method. Specifically, we target at answering the following questions in the experiments: 
\begin{itemize}[leftmargin=*]
\item What is the impact of BFI on the performance of cross networks?
\item How does the shuffling operation affect the performance?
\item What is the effect of feature interaction weight sharing?
\end{itemize}

\textbf{Datasets.}
We show the experiments on two public datasets: Covertype\footnote{http://archive.ics.uci.edu/dataset/31/covertype} with 54 features, 290,506 training instances and 290,506 testing instances; and YouTube Multiview Video Games Dataset\footnote{https://archive.ics.uci.edu/dataset/269/youtube+multiview+video+games+dataset} (YSpectro) having 1,024 features, 97,934 training data and 11,930 testing data.


\begin{figure*}[htbp]
\includegraphics[width=.24\textwidth]{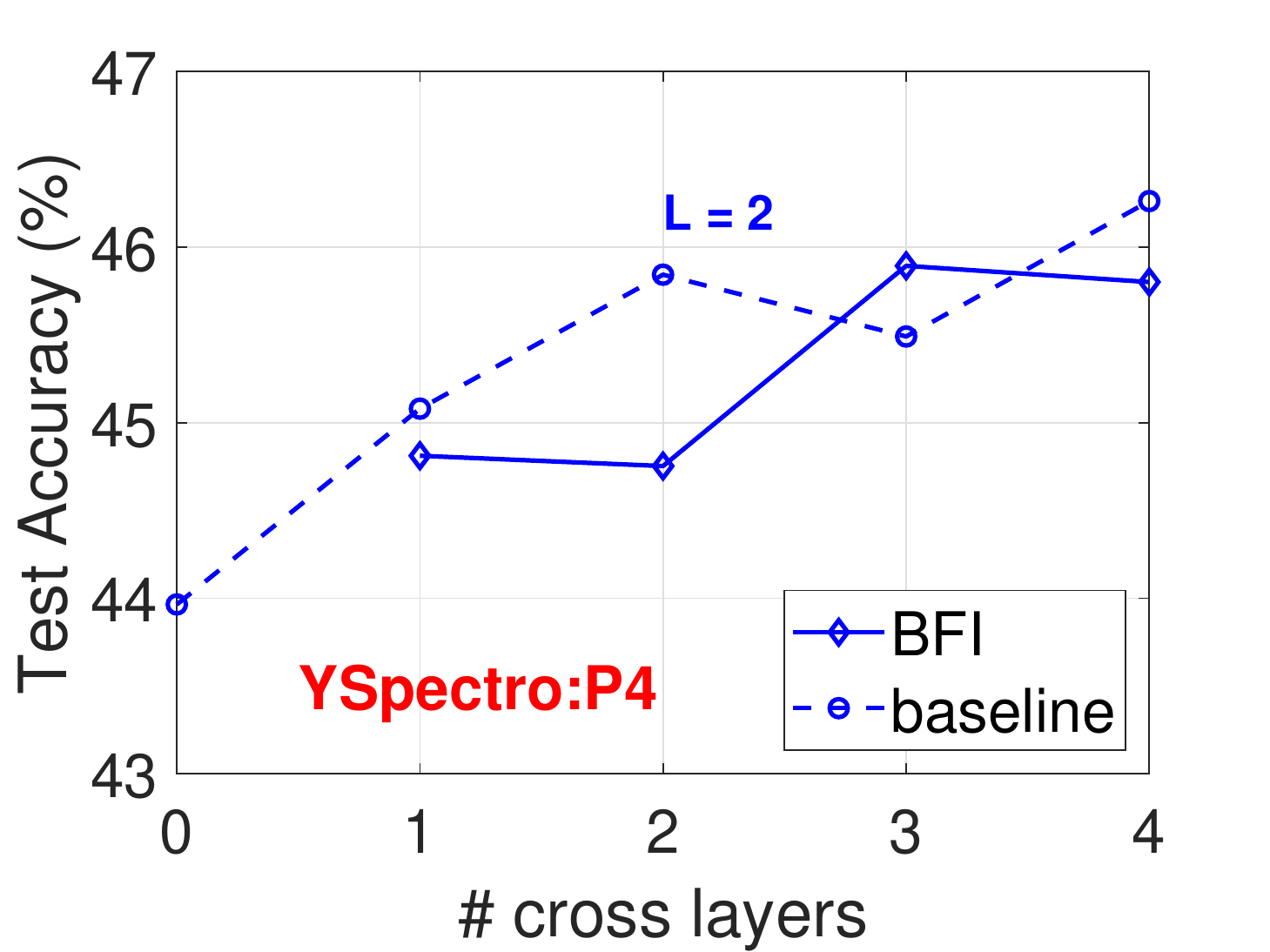}
\includegraphics[width=.24\textwidth]{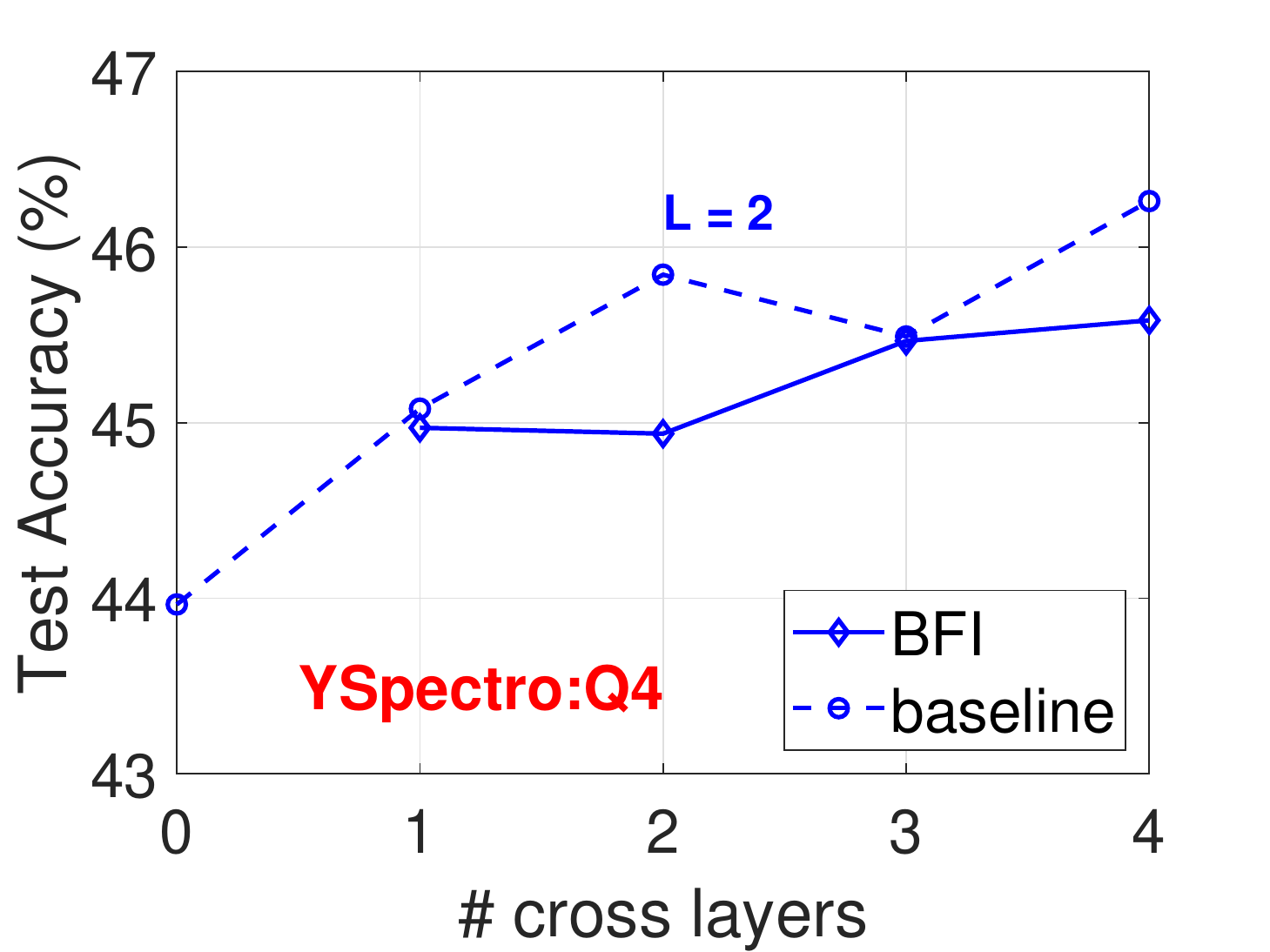}
\includegraphics[width=.24\textwidth]{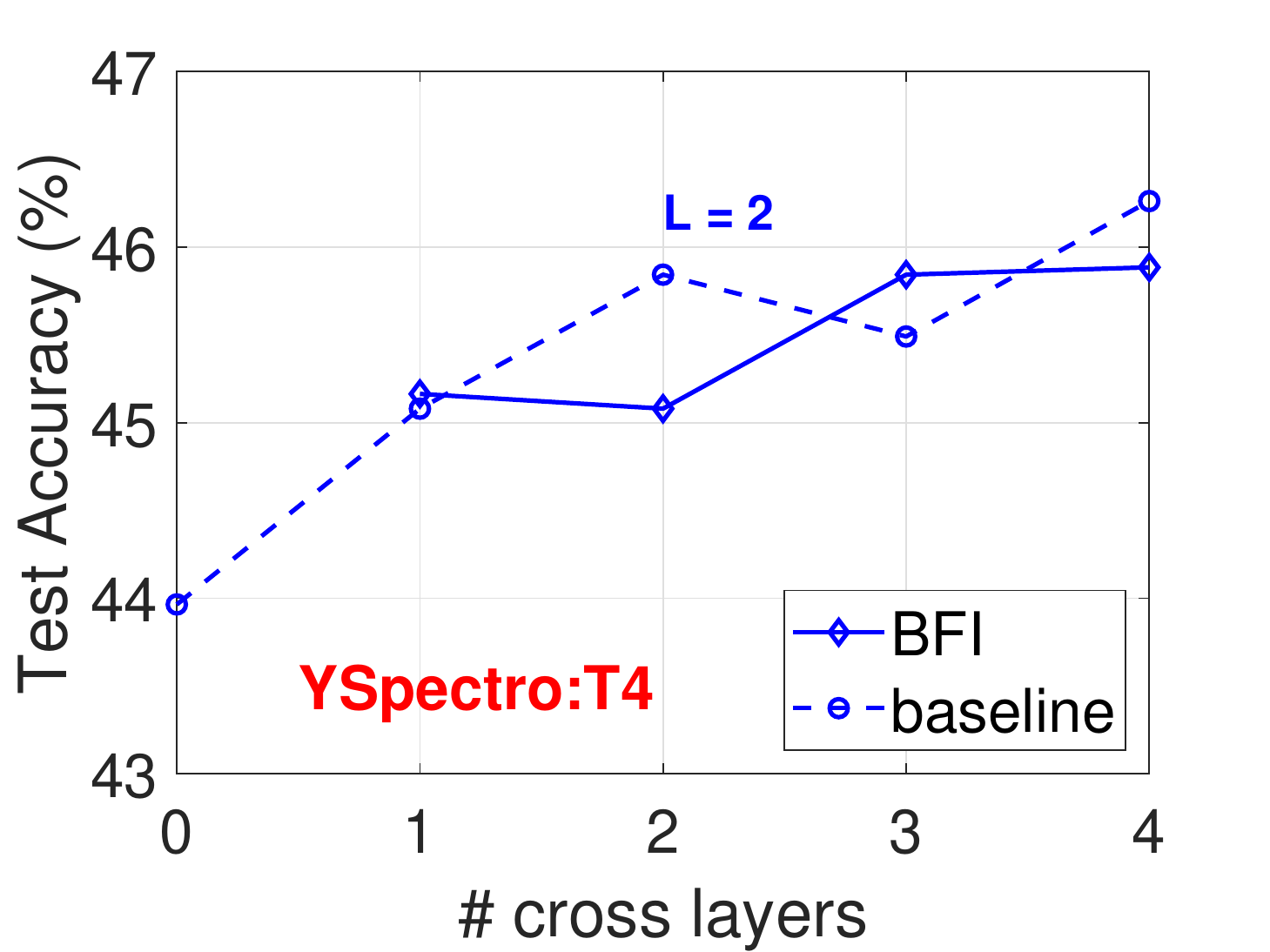}
\includegraphics[width=.24\textwidth]{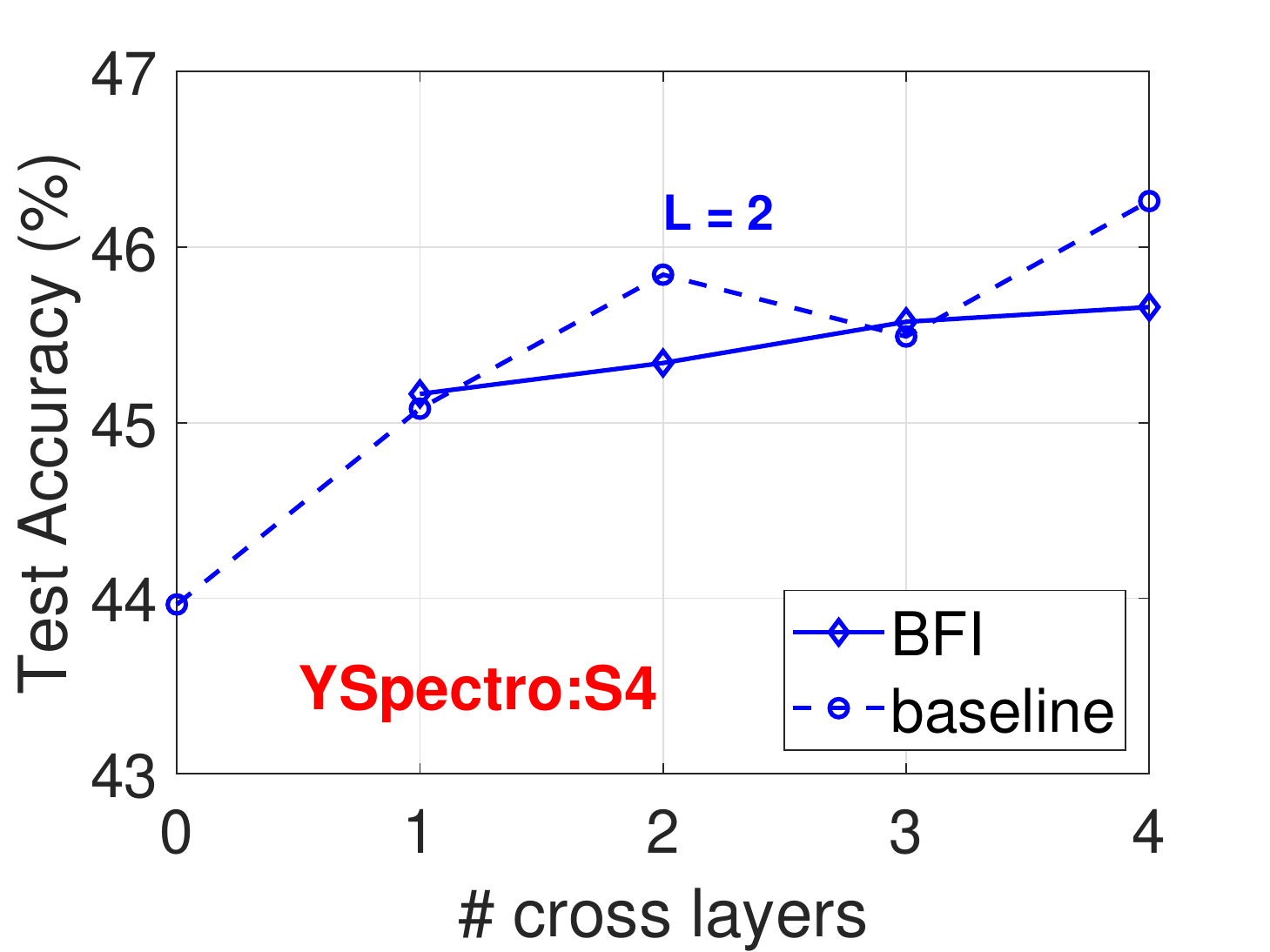}\\
\includegraphics[width=.24\textwidth]{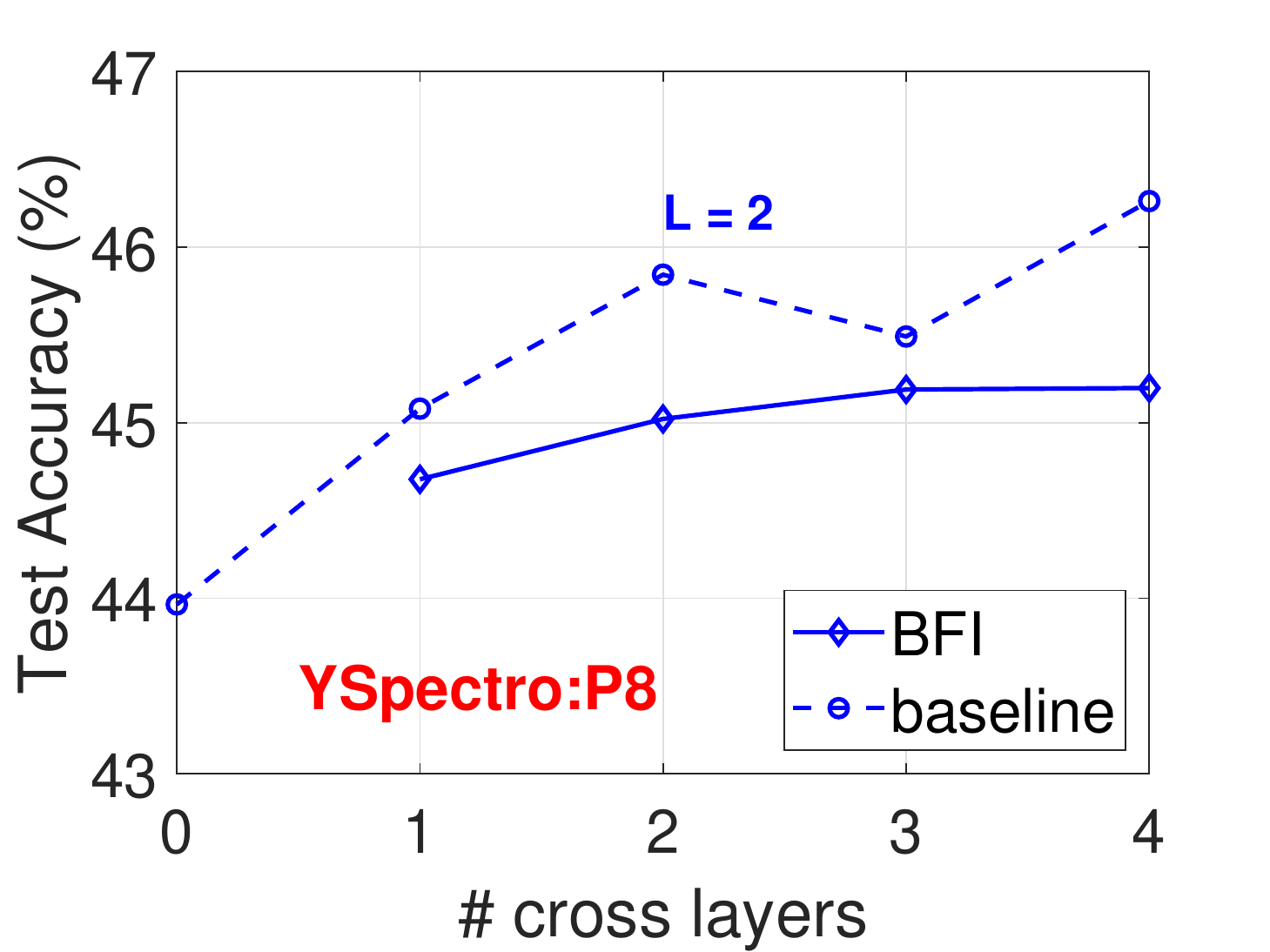}
\includegraphics[width=.24\textwidth]{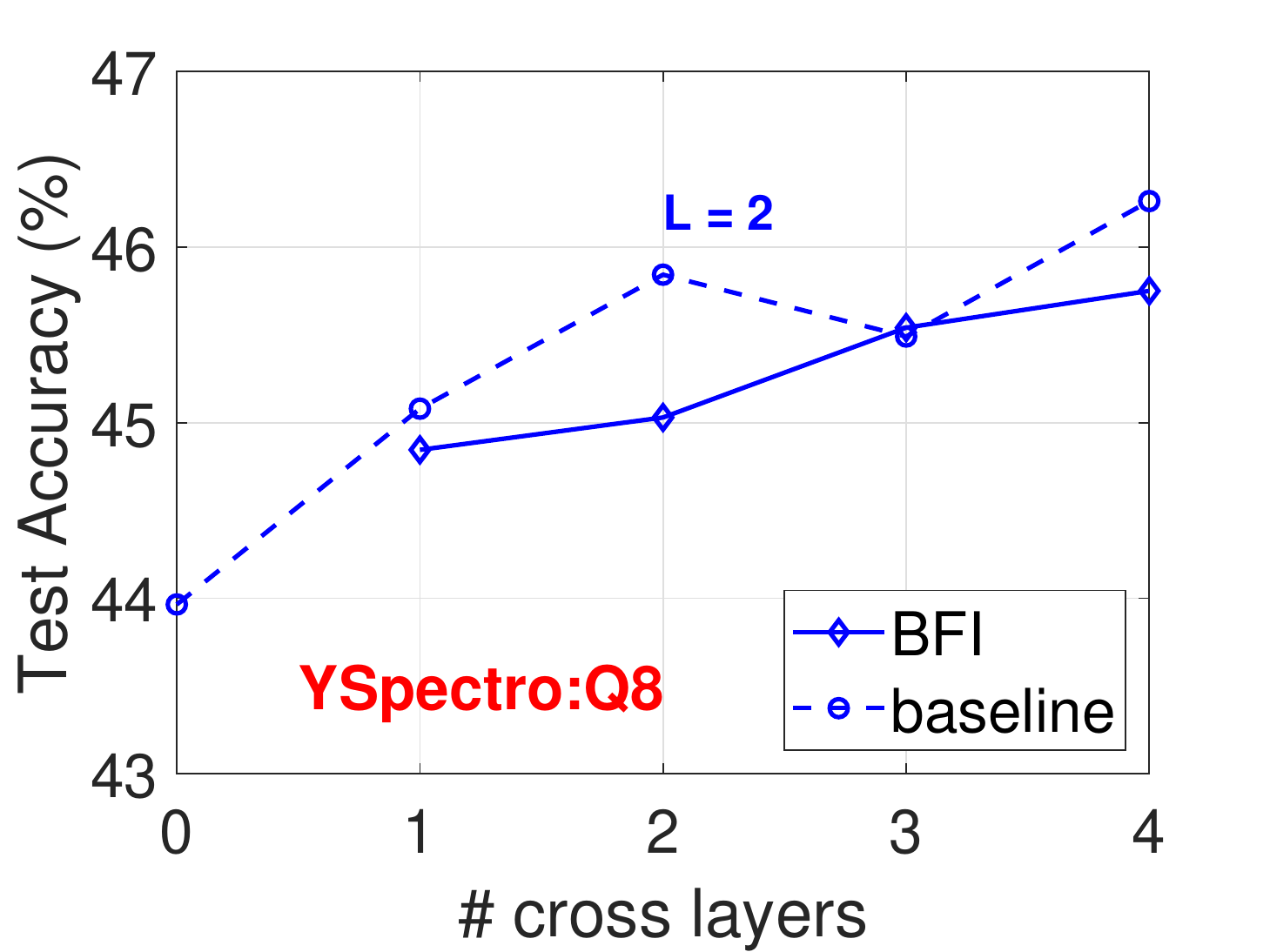}
\includegraphics[width=.24\textwidth]{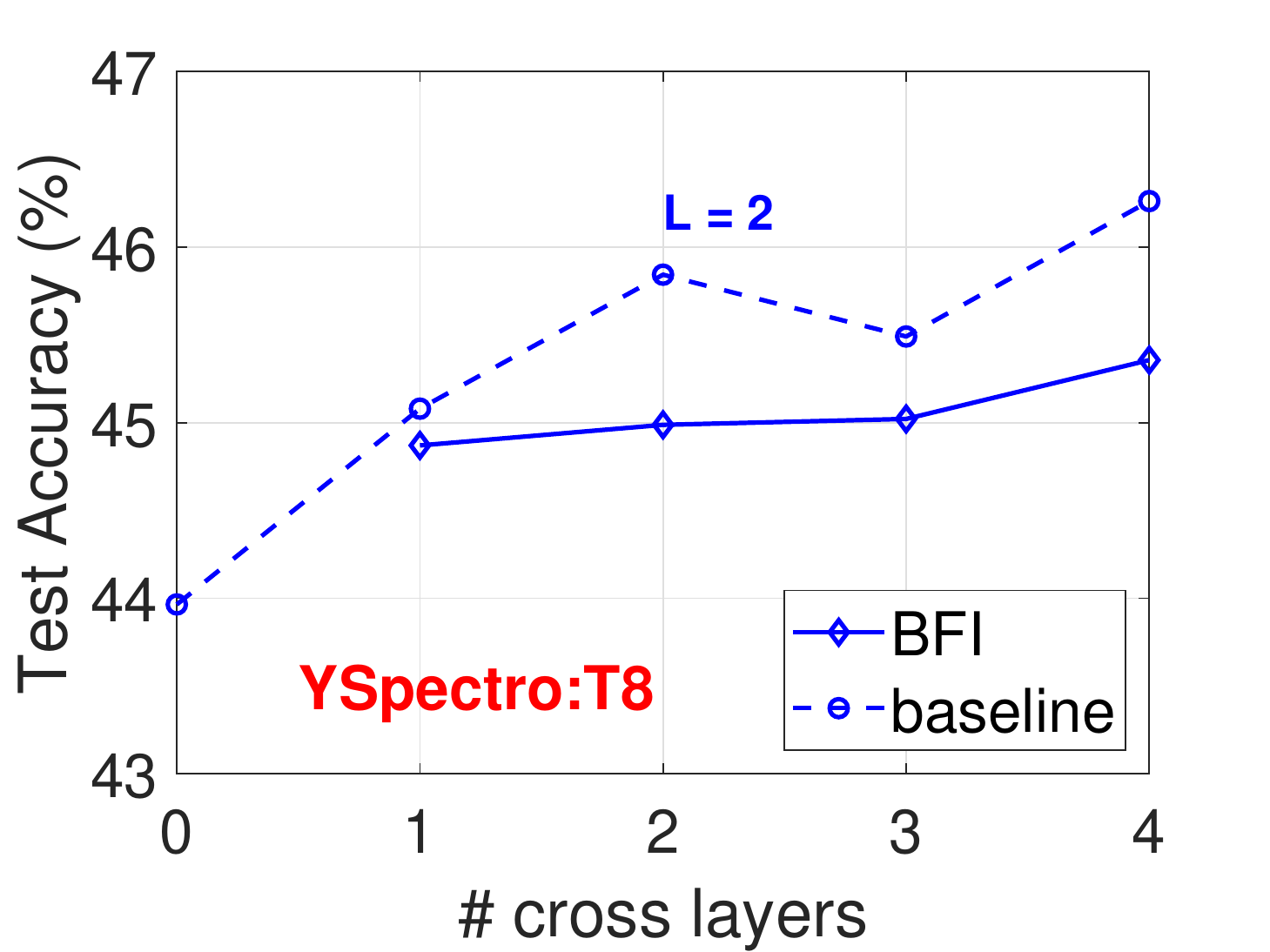}
\includegraphics[width=.24\textwidth]{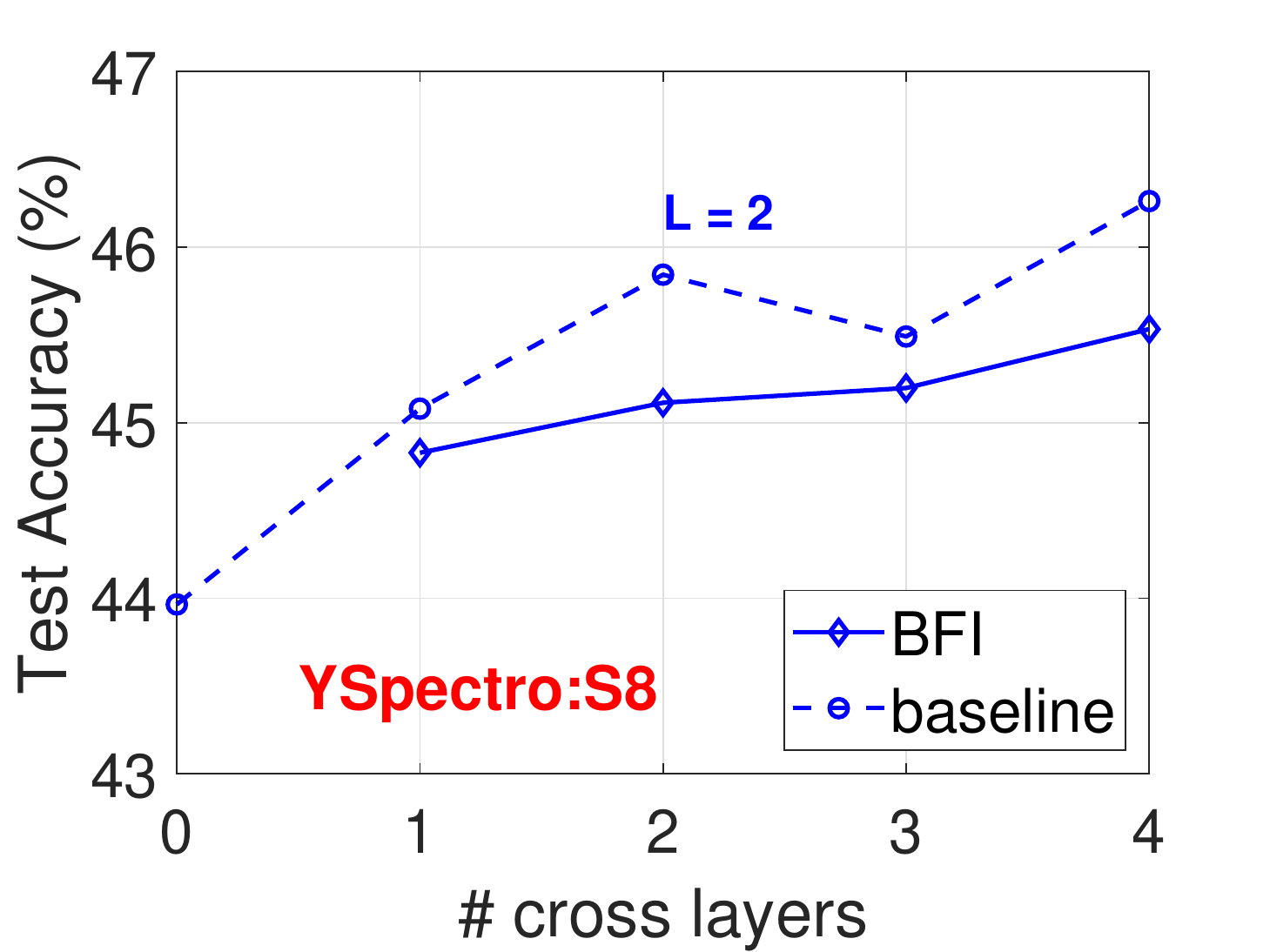}
\caption{Test accuracy for BFI and the cross network in DCNv2 (baseline) on YSpectro.}\label{fig:spectro}
\end{figure*}

\textbf{Models.} Classification is the task of our model. The raw features first go through the cross network/BFI. Then a given number ($L$) of fully-connected layers (256 neurons) are followed. The last layer is a fully-connected layer that regresses to the number of classes. The activation function we used is ReLU~\cite{brownlee2019gentle}. We implemented the network of BFI in PyTorch~\cite{paszke2017automatic} 1.13.1, Python 3.8.15.

Figure~\ref{fig:covtype} and Figure~\ref{fig:spectro} show the test accuracy of BFI and the baseline (DCNv2). We tested 4 variants of BFI denoted with P, Q, T, and S, respectively (Table~\ref{tbl:legend}). For block factor $K$, we use 3, 6, 9 on Covertype and 4, 8 on YSpectro. We only report the case when the number of fully-connected layers is 2 ($L=2$) for YSpectro because 1, 2, and 3 show very similar trends. We only keep case 2 for a cleaner view.

\textbf{Performance of BFI.}
P is the case when we pay the highest cost in BFI---we enable shuffling on every cross layer and disable weight sharing. For Covertype, the difference between BFI and baseline is minimum when block factor $K$ is 3 (BFI only requires $1/3$ computation and memory on cross network than the baseline). As expected, when we use a greater $K$, the gap becomes larger. 

On the other hand, note that using more cross layers captures higher order of feature interaction and potentially generates better results. When we compare the performance of BFI and the baseline with the same cost, e.g., the baseline uses 1 cross layer while BFI uses 3 cross layers on Covertype, BFI shows a higher test accuracy. This trend is also true on YSpectro: given a fixed computation and memory budget, using BFI enables us to have more cross layers and yield better performance.

\textbf{Effect of shuffling.}
To analyze the effect of shuffling (whether the cross block interaction matters), we compare P with Q, and T with S. The gap is obvious when the number of following fully-connected layers is 1. However, the gap shrinks when we have more layers. The reason is that the latter fully-connected layers can also exploit the feature interactions. Although the higher order interactions are not captured across blocks in the cross networks, those latter fully-connected layers capture them. 

\textbf{Weight sharing.}
We compare P with T and Q with S to explore the effect of weight sharing. The sharing almost always degrades the performance for both datasets. However, the performance gap becomes negligible when we have a sufficient number of following fully-connected layers. Note that the weight sharing reduces the memory footprint of cross networks by a factor of $K$. We can conclude that the weight sharing is effective especially when we can fit the network into GPU memory with this memory saving or the following layer is sufficiently deep.

\vspace{.1in}
\section{Conclusions}
Feature interactions play a crucial role in uncovering complex dependencies and improving the accuracy of recommendations. This paper presented the Blockwise Feature Interaction Network (BFI) as an approach for capturing feature interactions. BFI reduces both computation and memory consumption of cross networks for modeling feature interactions. We propose 4 variants of BFI and analyze the effect of shuffling and weight sharing. We experimentally evaluate the proposed methods on public datasets. The results demonstrated that the BFI method effectively captures feature interactions and improves the cross network performance while having the same computation and memory budget. Furthermore, by sharing embeddings, we achieved substantial reductions in memory usage without sacrificing much model effectiveness. Weight sharing proved to be a viable strategy for reducing memory footprint, which is especially crucial for resource-constrained environments such as GPU-based training and inference. We hope this work can provide valuable insights for using feature interactions. 


\balance
\bibliographystyle{plain}
\bibliography{refs_scholar}

\end{document}